\documentclass[aps,pra,reprint, onecolumn, citeautoscript, nofootinbib, superscriptaddress]{revtex4-2}
\setcitestyle{super}

\usepackage[T1]{fontenc}
\usepackage{lmodern}
\usepackage[utf8]{inputenc}
\usepackage{amsmath,amssymb}
\usepackage{graphicx}
\usepackage{hyperref}
\usepackage{geometry}
\usepackage{color}
\date{May 2026}

\begin{document}
\title{PIQC: Scalable Distributed Quantum Computing via Photonic Integration of Designed Molecular Quantum Nodes}

\author{Anna Aubele}
\affiliation{NVision Quantum Technologies GmbH, Wolfgang-Paul-Str. 2, 89081 Ulm, Germany}

\author{Gregor Bayer}
\affiliation{NVision Quantum Technologies GmbH, Wolfgang-Paul-Str. 2, 89081 Ulm, Germany}

\author{Tim R. Eichhorn}
\affiliation{NVision Quantum Technologies GmbH, Wolfgang-Paul-Str. 2, 89081 Ulm, Germany}

\author{Tobias Hahn}
\affiliation{NVision Quantum Technologies GmbH, Wolfgang-Paul-Str. 2, 89081 Ulm, Germany}

\author{Fedor Jelezko}
\affiliation{NVision Quantum Technologies GmbH, Wolfgang-Paul-Str. 2, 89081 Ulm, Germany}
\affiliation{Institute for Quantum Optics (IQO), Ulm University, Albert-Einstein-Allee 11, 89081 Ulm, Germany}
\affiliation{Center for Integrated Quantum Science and Technology (IQST), Ulm University, 89081 Ulm, Germany}

\author{Paul Mentzel}
\affiliation{NVision Quantum Technologies GmbH, Wolfgang-Paul-Str. 2, 89081 Ulm, Germany}

\author{Philipp Neumann}
\affiliation{NVision Quantum Technologies GmbH, Wolfgang-Paul-Str. 2, 89081 Ulm, Germany}

\author{Matthias Pfender}
\affiliation{NVision Quantum Technologies GmbH, Wolfgang-Paul-Str. 2, 89081 Ulm, Germany}

\author{Martin B. Plenio}
\affiliation{NVision Quantum Technologies GmbH, Wolfgang-Paul-Str. 2, 89081 Ulm, Germany}
\affiliation{Institute of Theoretical Physics, Ulm University, Albert Einstein Allee 11, 89081 Ulm, Germany}
\affiliation{Center for Integrated Quantum Science and Technology (IQST), Ulm University, 89081 Ulm, Germany}

\author{Alex Retzker}
\affiliation{Racah Institute of Physics, The Hebrew University of Jerusalem, Jerusalem 9190401, Israel}

\author{Simon Roggors}
\affiliation{NVision Quantum Technologies GmbH, Wolfgang-Paul-Str. 2, 89081 Ulm, Germany}
\affiliation{Institute for Quantum Optics (IQO), Ulm University, Albert-Einstein-Allee 11, 89081 Ulm, Germany}

\author{Alon Salhov}
\affiliation{NVision Quantum Technologies GmbH, Wolfgang-Paul-Str. 2, 89081 Ulm, Germany}

\author{Jochen Scharpf}
\affiliation{NVision Quantum Technologies GmbH, Wolfgang-Paul-Str. 2, 89081 Ulm, Germany}

\author{Tobias A. Schaub}
\affiliation{NVision Quantum Technologies GmbH, Wolfgang-Paul-Str. 2, 89081 Ulm, Germany}

\author{Nico Striegler}
\affiliation{NVision Quantum Technologies GmbH, Wolfgang-Paul-Str. 2, 89081 Ulm, Germany}
\affiliation{Institute for Quantum Optics (IQO), Ulm University, Albert-Einstein-Allee 11, 89081 Ulm, Germany}

\author{Thomas Unden}
\affiliation{NVision Quantum Technologies GmbH, Wolfgang-Paul-Str. 2, 89081 Ulm, Germany}

\author{Julia Zolg}
\affiliation{NVision Quantum Technologies GmbH, Wolfgang-Paul-Str. 2, 89081 Ulm, Germany}
\affiliation{Institute of Organic Chemistry III, Ulm University, Albert-Einstein-Allee 11, 89081 Ulm, Germany}
\affiliation{Center for Integrated Quantum Science and Technology (IQST), Ulm University, 89081 Ulm, Germany}

\author{Sella Brosh}
\affiliation{NVision Quantum Technologies GmbH, Wolfgang-Paul-Str. 2, 89081 Ulm, Germany}

\author{Ilai Schwartz}
\email{ilai@nvision-imaging.com}
\affiliation{NVision Quantum Technologies GmbH, Wolfgang-Paul-Str. 2, 89081 Ulm, Germany}

\begin{abstract}
There is a growing consensus that large-scale, fault-tolerant quantum computing (FTQC) necessitates high-fidelity photonic interconnects to overcome the scaling limits of monolithic architectures \cite{bourassaEntanglementControlSingle2020a,barralReviewDistributedQuantum2025, monroeLargescaleModularQuantumcomputer2014, awschalomDevelopmentQuantumInterconnects2021}. However, most current platforms were not originally designed for native photonic connectivity and require significant engineering overhead \cite{knautEntanglementNanophotonicQuantum2024,humphreysDeterministicDeliveryRemote2018a}. To overcome these fundamental hardware limitations, we recently introduced a rationally designed organic molecule that serves as an ideal quantum node, featuring a robust qubit-photon interface (QPI)\cite{roggorsSingleMoleculeSpinPhotonInterface2026} and a long-lived nuclear-spin register.

In this work, we present \textbf{PIQC (Photonic Integrated Quantum Circuits)}, a distributed architecture designed to scale these molecular nodes into a functional quantum computer\cite{nickersonFreelyScalableQuantum2014, nemotoPhotonicArchitectureScalable2014a}. 

The PIQC framework integrates five mutually reinforcing innovations: (i) \textbf{Designer molecular qubits}, \textit{i.e.} carbene molecules in an isosteric host that provide millisecond-coherence electron spins with high spectral stability and spin-dependent optical emission\cite{roggorsOpticallyDetectedMagnetic2025a,roggorsSingleMoleculeSpinPhotonInterface2026}, (ii) \textbf{deterministic nuclear registers} made of synthetically placed $^{13}$C or $^{14}$N labels that enable fast ($\sim 1~\mu$s), high-fidelity electron-nuclear gates\cite{waldherrQuantumErrorCorrection2014a,taminiauUniversalControlError2014,bradleyTenQubitSolidStateSpin2019a}, (iii) \textbf{hybrid photonic integration}, which allows molecular films to seamlessly integrate with existing mature fabrication technologies, \textit{e.g.} thin-film lithium niobate (TFLN),~\cite{Huetal2024} (iv) \textbf{heralded entanglement protocols} that can tolerate up to 70\% photon loss\cite{barrettEfficientHighfidelityQuantum2005a,wangIntegratedLithiumNiobate2018,lombardiTriggeredEmissionIndistinguishable2021}, and (v) \textbf{stairway Floquetification,} \textit{i.e.} high-rate quantum low-density parity-check (qLDPC) codes that are converted into Floquet codes, reducing syndrome extraction to weight-two Bell-pair measurements that match PIQC's networked hardware\cite{hastingsDynamicallyGeneratedLogical2021,gidneyFaultTolerantHoneycombMemory2021,vuillotPlanarFloquetCodes2021,brownFaulttolerantNonCliffordGate2020,jacobyStairwayCodesFloquetifying2026}.

PIQC offers a hardware-efficient, commercially viable pathway toward a utility-scale quantum computer based on distributed FTQC\cite{gidneyHowFactor20482025,gidneyHowFactor20482021a,gouzienPerformanceAnalysisRepetition2023}.
\end{abstract}

\maketitle

\section{Introduction}
Driven by substantial financial investments and growing public awareness, the quantum computing sector is currently undergoing a period of unprecedented expansion and accelerated technological development.
This progress is evidenced by the significant milestones recently achieved across various quantum computing platforms, indicating a transition from the noisy intermediate-scale quantum (NISQ) era toward fault-tolerant quantum computing (FTQC).
Trapped ions, superconducting qubits, and neutral-atom arrays have emerged as leading technologies, with demonstrations of below-threshold fault-tolerant operation on hundreds of qubits \cite{bluvsteinFaulttolerantNeutralatomArchitecture2026,acharyaQuantumErrorCorrection2025,reichardtDemonstrationQuantumComputation2024}.

In parallel, improvements in quantum error correction (QEC) architectures, especially using high-rate qLDPC codes\cite{postolProposedQuantumLow2001,breuckmannQuantumLowDensityParityCheck2021}, have enabled previously unimaginable ratios of 50\% logical to physical qubits \cite{breuckmannQuantumLowDensityParityCheck2021,panteleevDegenerateQuantumLDPC2021,bravyiHighthresholdLowoverheadFaulttolerant2024}. Leveraging these advancements, recent resource estimates show that Shor’s algorithm may be executable with as few as 10\,000 qubits \cite{cainShorsAlgorithmPossible2026}, bringing us very close to a universal quantum computer with a broad range of applications.

However, all current QC architectures are based on single, ``monolithic'' chips. We have seen that the challenge of scaling these chip architectures by several orders of magnitude remains insurmountable despite the massive investment and focus by leading companies and research institutes worldwide. For this reason, there is a growing consensus in the community that photonic interconnectivity is necessary for building quantum computers with hundreds of thousands or more qubits.

Building the architecture for distributed fault-tolerant quantum computing requires a unique interface between photons and qubits. 
This QPI remains the critical bottleneck for realizing distributed quantum computing; despite substantial research efforts, an interface that simultaneously achieves the necessary scalability, high fidelity, and high rate has yet to be demonstrated. The challenge is formidable: The system needs to simultaneously provide high-fidelity local qubit control coupled with an efficient, high-rate, quantum-state-dependent optical interface for remote entanglement generation. 

Moreover, compatibility with existing fabrication processes provides a critical advantage for integrating tens of thousands to millions of QPIs required for fault-tolerant quantum computing.

When enumerating all the requirements for these QPIs, one arrives at a challenging list of criteria, including\footnote{The current QPI requirements could vary for alternative approaches, for example for single-photon cavity reflection schemes.}:
\begin{itemize}
    \item Spin-dependent optical emission, for high-fidelity qubit-photon entanglement
    \item High-rate photon emission, including short excited-state decay time, high emission to the ZPL and high Debye-Waller factor.
    \item Narrow, stable optical linewidth, capable of emitting indistinguishable photons from different molecules.
    \item Long coherence time (>1 ms) for enabling high fidelity operations with the qubits.
\end{itemize}
A variety of physical platforms have been fostered, including trapped ions and semiconductor quantum dots.
While trapped ions have recently been utilized to demonstrate deterministic quantum gate teleportation and distributed quantum computing across an optical network link\cite{mainDistributedQuantumComputing2025}, quantum dots provide deterministic, high-rate single-photon emission, serving as the foundation for practical, low-depth fault-tolerant photonic quantum computing architectures\cite{larsenDeterministicGenerationTwodimensional2019,zhaiQuantumInterferenceIdentical2022,chanPracticalBlueprintLowdepth2025}.
Alongside these platforms, solid-state defect centers, most prominently silicon-vacancy (SiV) \cite{knautEntanglementNanophotonicQuantum2024, riedelScalablePhotonicQuantum2026} or nitrogen-vacancy\cite{iulianoUnconditionallyTeleportedQuantum2026} centers in diamond, T centers in silicon\cite{incDistributedQuantumComputing2024a} and V2 centers in silicon carbide\cite{sonDevelopingSiliconCarbide2020} have demonstrated many of the required criteria for a QPI when selecting a nuclear spin as the qubit.
To date, no leading platform has simultaneously fulfilled all criteria, as each remains constrained by at least one critical limitation.
In the case of existing solid-state defects, we believe this hurdle is fundamental - researchers are strictly bound by the material's intrinsic properties, leaving virtually no room to tune a quantum color center if it fails to meet the exact needs of an ideal QPI.\cite{toganQuantumEntanglementOptical2010,nguyenQuantumNetworkNodes2019a,bhaskarExperimentalDemonstrationMemoryenhanced2020,knautEntanglementNanophotonicQuantum2024,higginbottomOpticalObservationSingle2022,sonDevelopingSiliconCarbide2020}

PIQC offers to solve this problem by leveraging the unique properties of organic carbene molecules as optically addressable spins.
Unlike atomic defects fixed in a covalent host lattice, molecular qubits can be tuned with atomic precision through organic synthesis: spin structure, optical transition wavelength, zero-field splitting (ZFS) parameters, and nuclear-spin labels as quantum memories are all accessible by rational molecular design.
In a recent breakthrough, we have established that a comparatively simple diarylcarbene molecule (BiPhi), embedded in an isosteric crystalline molecular host, functions as a robust spin-QPI with millisecond electron spin coherence, spin-selective optical transitions, narrow optical linewidth and exceptional spectral stability, demonstrating for the first time that the requirements for a QPI can be comprehensively met by a purely organic molecular system\cite{roggorsSingleMoleculeSpinPhotonInterface2026}.

PIQC integrates these molecular qubits onto TFLN photonic circuits that provide tuneable resonator-enhanced emission, electro-optic switching, Bell-state measurement (BSM) stations, coupled with control electronics for fast and high-fidelity local control of electron-nuclear spin registers.
For fault tolerance, we apply the Stairway Floquetification procedure\cite{jacobyStairwayCodesFloquetifying2026} to high-rate quantum LDPC (qLDPC) codes.
This converts each weight-$w$ stabilizer into a periodic sequence of weight-two measurements, each implementable by a single photonic Bell pair, which is a natural match to PIQC's distributed architecture.
Applying the Floquetification procedure to the most recent qLDPC codes\cite{cainShorsAlgorithmPossible2026,babbushSecuringEllipticCurve2026} enables us to achieve exceptional logical to physical qubit ratios.

This perspective paper is organized as follows.
Section \ref{sec:biphi} describes the BiPhi molecular qubit.
Section \ref{sec:spi} covers the spin--photon interface and TFLN photonic integration.
Section \ref{sec:sysarch} presents the full system architecture.
Section \ref{sec:distributed_qec} introduces the Stairway Floquetification of high-rate qLDPC codes. Section \ref{sec:conclusion} concludes.

\section{The PIQC Molecular Qubit}
\label{sec:biphi}
\subsection{Molecular design: The BiPhi diarylcarbene}
\begin{figure}[h]
\centering
\includegraphics[width=0.75\textwidth]{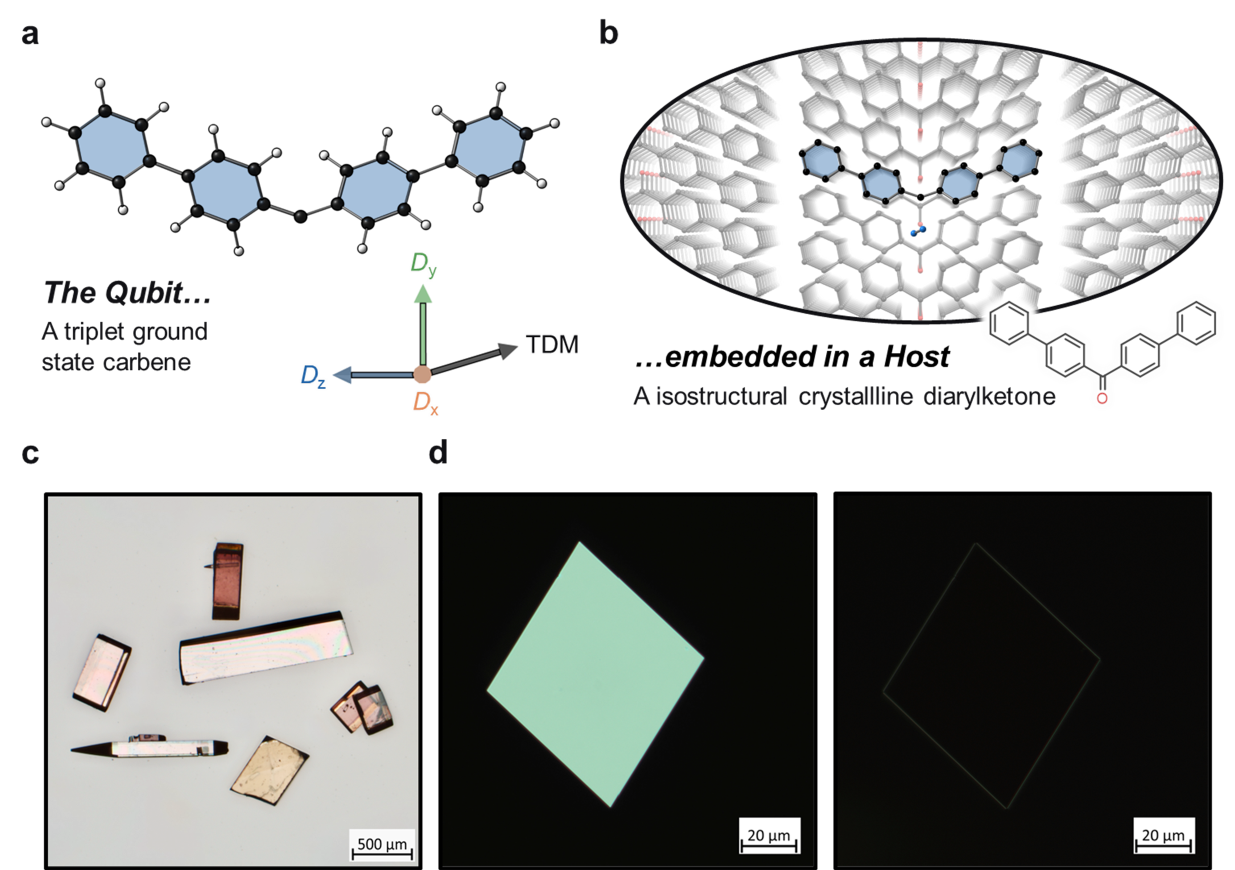}
\caption{\textbf{a}, Molecular structure of the qubit together with the orientation of the ZFS tensor ($D_i$) and optical transition dipole moment (TDM)\cite{roggorsSingleMoleculeSpinPhotonInterface2026}. \textbf{b}, Illustration of the embedding of the carbene and the nitrogen cleaved off via photoactivation of a diaryldiazomethane precursor in the crystalline lattice of ketone matrix\cite{roggorsSingleMoleculeSpinPhotonInterface2026}. 
\textbf{c}, Representative crystals with a uniform 2.5 mol\% doping rate shown in multiple orientations. Different colors stem from variations of the effective optical density along the viewing axis, scaling with crystal thickness and orientation\cite{roggorsSingleMoleculeSpinPhotonInterface2026}. \textbf{d}, Polarization microscopy images of a submicron-thin crystal. Left: The crystal viewed at maximum brightness. Right: The crystal viewed with crossed polarizers rotated 45° from the bright orientation, making it appear completely dark except for light scattering at the edges and indicating a high degree of uniform crystallinity.} 
\label{fig:1}
\end{figure}
The PIQC qubit material is composed of a photoactivated di([1,1'-biphenyl]-4-yl)carbene (Fig.~\ref{fig:1}a), referred to as BiPhi, embedded in an isosteric diarylketone crystalline host ((Fig.~\ref{fig:1}b).
BiPhi is a bent, pseudo-$C_2$-symmetric diarylcarbene consisting of two \textit{para}-connected biphenyls which converge at a divalent carbene carbon.
Two strongly correlated unpaired electrons give rise to a spin-1 ($S=1$) triplet ground state $T_0$ \cite{roggorsSingleMoleculeSpinPhotonInterface2026}.

The BiPhi molecule exhibits optical transitions in the red spectral region around 600 nm, where it features both a triplet ground state ($T_0$) and a triplet excited-state ($T_1$). The ground-state ZFS parameters are $|D| = 11\,160$ MHz (axial) and $|E| = 541$ MHz (rhombic), shifting to $D \approx -5\,110$ MHz and $E \approx 240$ MHz in the excited-state\cite{roggorsSingleMoleculeSpinPhotonInterface2026}. The large detuning of $\approx 17$ GHz between the light and dark spin states, compared to the narrow linewidth ($\sim 38$~MHz) is key to spin-selective photon emission and efficient optical readout of spin states.

A distinctive advantage of BiPhi as a qubit platform lies in the material properties of the host system. The host matrix is an isosteric diarylketone, a structurally related molecule where the carbene carbon is replaced by a carbonyl group. This isostericity allows the BiPhi molecules to  seamlessly substitute into the host lattice at controlled doping concentrations (typically mol\% - ppb) with minimal strain or disorder. High-quality crystalline thin films can be grown by physical vapor deposition, from solution, or other standard solution processing techniques that are scalable and inexpensive compared to semiconductor crystal growth \cite{roggorsSingleMoleculeSpinPhotonInterface2026,wasielewskiExploitingChemistryMolecular2020b,baylissOpticallyAddressableMolecular2020c,baylissEnhancingSpinCoherence2022a}. The crystalline host environment also confers important photophysical advantages in that it planarizes the biphenyl units of BiPhi (Ph--Ph dihedral angle $\approx 6^\circ$ vs.\ $36^\circ$ in the gas phase). The constrained geometry suppresses non-radiative decay channels by reducing the Franck--Condon factor for phonon coupling which results in a high photoemission rate into the zero-phonon line (ZPL) of 20\%, which can be increased further by coupling to an optical cavity.

Crucially, the molecular qubit can be chemically modified: Isotopic labelling with $^{13}$C, $^{14}$N, $^{19}$F, $^{31}$P or other isotopes, at specific positions engineers the nuclear spin register, while modifications to the biphenyl substituents can tune the optical transition wavelength and ZFS parameters. Navigating the interdependencies of these two design principles is where the full power of computational predictions and synthetic chemistry comes into play, ultimately providing a level of design freedom unavailable in solid-state defect platforms\cite{roggorsSingleMoleculeSpinPhotonInterface2026,wasielewskiExploitingChemistryMolecular2020b}.

\subsection{Single-molecule spin-photon interface}
\begin{figure}[h]
\centering
\includegraphics[width=\textwidth]{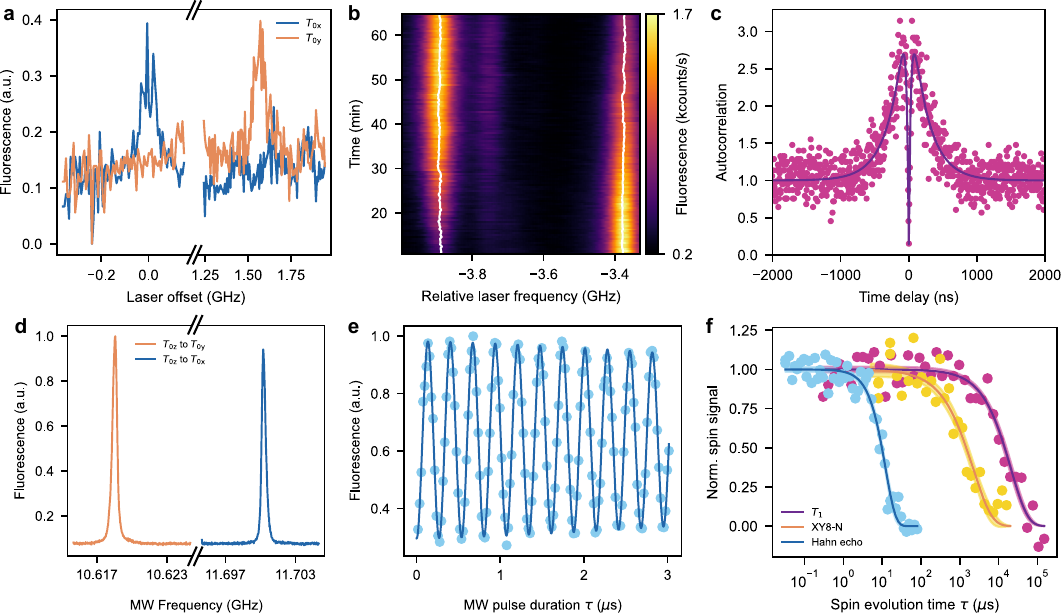}
\caption{\textbf{a}, Excitation spectra of a single BiPhi molecule depending on the internal spin state. Each spin state has a different excitation frequency (and produces a different photon frequency), in this case separated by 1.563\,GHz, thus enabling generation of photons entangled to the spin state. \textbf{b}, Longterm, repeated acquisition of fluorescence excitation spectra demonstrates spectral stability of the individual emitters with a standard deviation of the center frequency of about 2.6 MHz over one hour\cite{roggorsSingleMoleculeSpinPhotonInterface2026}. \textbf{c}, Fluorescence autocorrelation measurement confirms the detection of a single emitter ($g^{(2)}(0) = 0.14$)\cite{roggorsSingleMoleculeSpinPhotonInterface2026}. \textbf{d}, The optically detected magnetic resonance (ODMR) spectrum demonstrating almost full on/off visibility.\cite{roggorsSingleMoleculeSpinPhotonInterface2026} \textbf{e}, Rabi oscillations of a single BiPhi molecule electron spin with a frequency of 3.7\,MHz. \textbf{f}, Lifetimes of a single BiPhi electron spin for various MW pulses sequences. During a Hahn-echo sequence, the spin signal decays after $12.2(6) \, \mu$s to $1/e$, during an XY8-N sequence (for N=1...2000) after 2.2(3)\,ms and during a $T_1$-sequence after 21(2)\,ms.}
\label{fig:2}
\end{figure}

Recent experimental work demonstrated the following key properties at $T = 4.5$~K, establishing BiPhi as a QPI\cite{roggorsSingleMoleculeSpinPhotonInterface2026}:

\begin{itemize}
    \item \textbf{Narrow, spin-dependent, optical emission:} resonant single-molecule excitation yields FWHM = 38 MHz already at 4.5\,K, which we expect to further improve at lower temperatures. The lifetime-limited linewidth is $\Delta\nu_{\text{lim}} = 6.6$\,MHz for $\tau_{\text{exc}} = 24$\,ns. As the difference between the optical excitation frequencies is up to 17~\,GHz, over 400-fold larger than the linewidth, this enables high fidelity spin-dependent emission. See Fig.~\ref{fig:2}a.
    \item \textbf{Spectral stability:} center-frequency fluctuations over $>1$~h remain at about 2.6\,MHz. See Fig.~\ref{fig:2}b.
    \item \textbf{Single-emitter addressability:} controlled \textit{in-situ} photoactivation resolves individual BiPhi molecules; photon autocorrelation confirms single-emitter emission with $g^{(2)}(0) = 0.14$. See Fig.~\ref{fig:2}c.
    \item \textbf{ODMR contrast:} Near-perfect single molecule ODMR contrast, enabling high-fidelity spin-state readout via optical cycling. See Fig.~\ref{fig:2}d.
    \item \textbf{Electron coherence:} XY8-N dynamical decoupling yields $T_2 = 2.2$\,ms, with $T_1 \approx 17$\,ms, even at the elevated temperature of 4.5\,K. We expect an additional improvement upon the deuteration of the dopant and host molecules. See Fig.~\ref{fig:2}f.
\end{itemize}

\subsection{Deterministic nuclear-spin qubits}
A key advantage of molecular qubits is the ability to engineer the local isotopic environment. PIQC incorporates nuclear-spin labels (\textit{e.g.} $^{13}$C or $^{14}$N) at designed molecular positions through chemical synthesis, yielding repeatable hyperfine coupling tensors in the few-to-tens-of-MHz range.
Thus, in PIQC we can use deterministically placed nuclear spins as memory/data qubits with high coherence time ($T_2^n > 10$\,ms) and with a strong coupling to the electron spin ($T_2^e > 2$\,ms), enabling fast gate times of $\sim 1 \, \mu$s.
At the same time, large ZFS parameters of the electron spin decouples both electron and nuclear spin from unwanted interaction\cite{miaoUniversalCoherenceProtection2020a}, further protecting the nuclear memory coherence during electron optical excitations~\cite{nvision_blueprint,cohen2017protecting}.
As the coherence time is 2\,000-fold larger than the gate time, 99.9\% electron-nuclear spin gate fidelity is achievable, crucial for fault-tolerant operation.\cite{nvision_blueprint,nvision_nuclear_electron_gates}

\section{Spin--Photon Interface and Silicon Photonic Integration}
\label{sec:spi}

Photons are the single quantum particle which can easily maintain coherence over long distances, leading to our conviction that all macroscale quantum connectivity necessitates photonic interconnects\cite{kimbleQuantumInternet2008,wehnerQuantumInternetVision2018,deboneThresholdsDistributedSurface2024,nickersonFreelyScalableQuantum2014,poulsennautrupFaulttolerantInterfaceQuantum2017a,iulianoUnconditionallyTeleportedQuantum2026,loblLosstolerantArchitectureQuantum2024}.

Photonic integrated circuits have significantly matured over the last ten years, largely propelled by the surging need for data-center interconnects and AI-accelerator networking.
The underlying investment wave has established reliable, high-yield foundry processes for low-loss waveguides, high-speed modulators, wavelength-division multiplexing filters, and on-chip photodetectors---all at wafer scale.
Crucially, the same fabrication infrastructure that produces 800G and 1.6T optical transceivers today can be repurposed for quantum photonic circuits with only incremental process modifications.

Beyond silicon photonics, TFLN has emerged as a complementary platform offering distinct advantages for quantum applications.
TFLN combines ultra-low propagation loss ($<0.2$\,dB/cm)~\cite{desiatovUltralowlossIntegratedVisible2019}, strong Pockels-effect electro-optic modulation with bandwidths exceeding 100\,GHz, and intrinsic second-order nonlinearity enabling on-chip frequency conversion and entangled photon-pair generation.
Recent demonstrations of large-scale hybrid integration---including quantum-dot emitters monolithically coupled to TFLN waveguide circuits---confirm the viability of heterogeneous photonic platforms for quantum networking\cite{zhangMonolithicUltrahighQLithium2017}.
The commercial TFLN ecosystem is expanding rapidly, with multiple foundries now offering multi-project wafer runs and design-kit support\cite{wangIntegratedLithiumNiobate2018,zhuIntegratedPhotonicsThinfilm2021,boesLithiumNiobatePhotonics2023,wangLargescaleQuantumDot2025}.

For PIQC, this convergence of photonic technology and industry investment is a strategic enabler. Rather than requiring bespoke quantum-only fabrication, our architecture leverages components (low-loss waveguides, high-speed switches, resonator filters, and fiber-chip couplers) that are being optimized and scaled by a multi-billion-dollar classical photonics industry.
The resulting cost curve, supply chain, and manufacturing maturity are assets that no other quantum modality can currently access at comparable scale, especially as in the PIQC architecture high photon loss (up to 70\%) can be tolerated.\cite{nvision_blueprint}

The PIQC molecules are built from first principles for being PIC compatible.
Our molecular thin films can be easily deposited on a variety of materials including silicon nitride and TFLN.
In particular, the latter offers mature low-loss waveguides, strong electro-optic modulation enabling nanosecond-scale switching, and compatibility with heterogeneous integration. 
Unlike covalent solid-state semiconductors, growing thin molecular crystals on top of PIC substrates eliminates common interface and lattice-matching issues arising from, e.g. epitaxial growth.
In addition, the carbene molecules are photo-activated from a reservoir of precursor molecules in the right density and at the right position, simplifying the integration with optical cavities on the PIC material. 

\section{System architecture}
\label{sec:sysarch}
\begin{figure}[h]
\centering
\includegraphics[width=\textwidth]{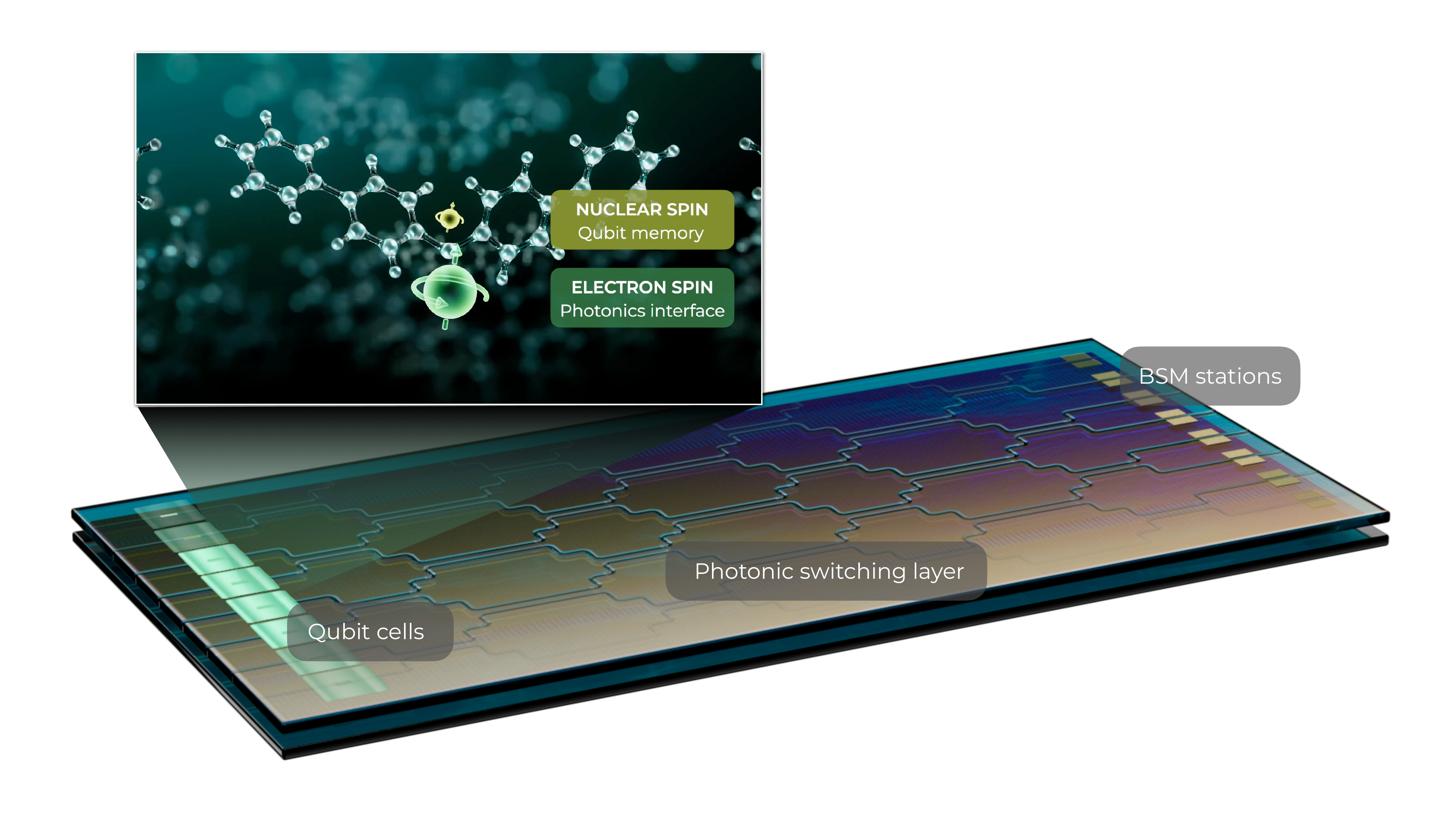}
\caption{Illustration of a PIQC chip, including the qubit cells, photonic switching layer and BSM stations.}
\label{fig:3}
\end{figure}

\subsection{Hardware layers}
The PIQC system decomposes into four hardware layers:
\begin{itemize}
    \item \textbf{Qubit cell} -- A qubit cell comprises a deposited thin molecular crystal film on a TFLN optical resonator. The molecular thin film includes BiPhi molecules with their QPIs and nuclear qubits, with one BiPhi molecule on resonance with the optical resonator. Both the molecule and the optical resonator are tuned electrically by electrodes to be on resonance, with each other and with the other qubit cells, building on hallmark demonstrations of precision electro-optic tuning in TFLN resonators \cite{zhangElectronicallyProgrammablePhotonic2019} and Stark-shift tuning of single organic molecules \cite{lettowQuantumInterferenceTunably2010}. The optical resonator is connected to an outcoupler waveguide.
    \item \textbf{Photonic switching layer} -- Reconfigurable switch fabric routing photons~\cite{psiquantum2025manufacturable, Huetal2024} from any qubit cell to any BSM, enabling long-range connectivity. The photonic switching layer can tolerate up to 70\% photon loss.
    \item \textbf{BSM stations} -- The optical detection. A BSM station includes a reconfigurable beam splitter, enabling the erasure of ``which path'' information or direct detection.
\end{itemize}

\begin{figure}[h]
\centering
\includegraphics[width=\textwidth]{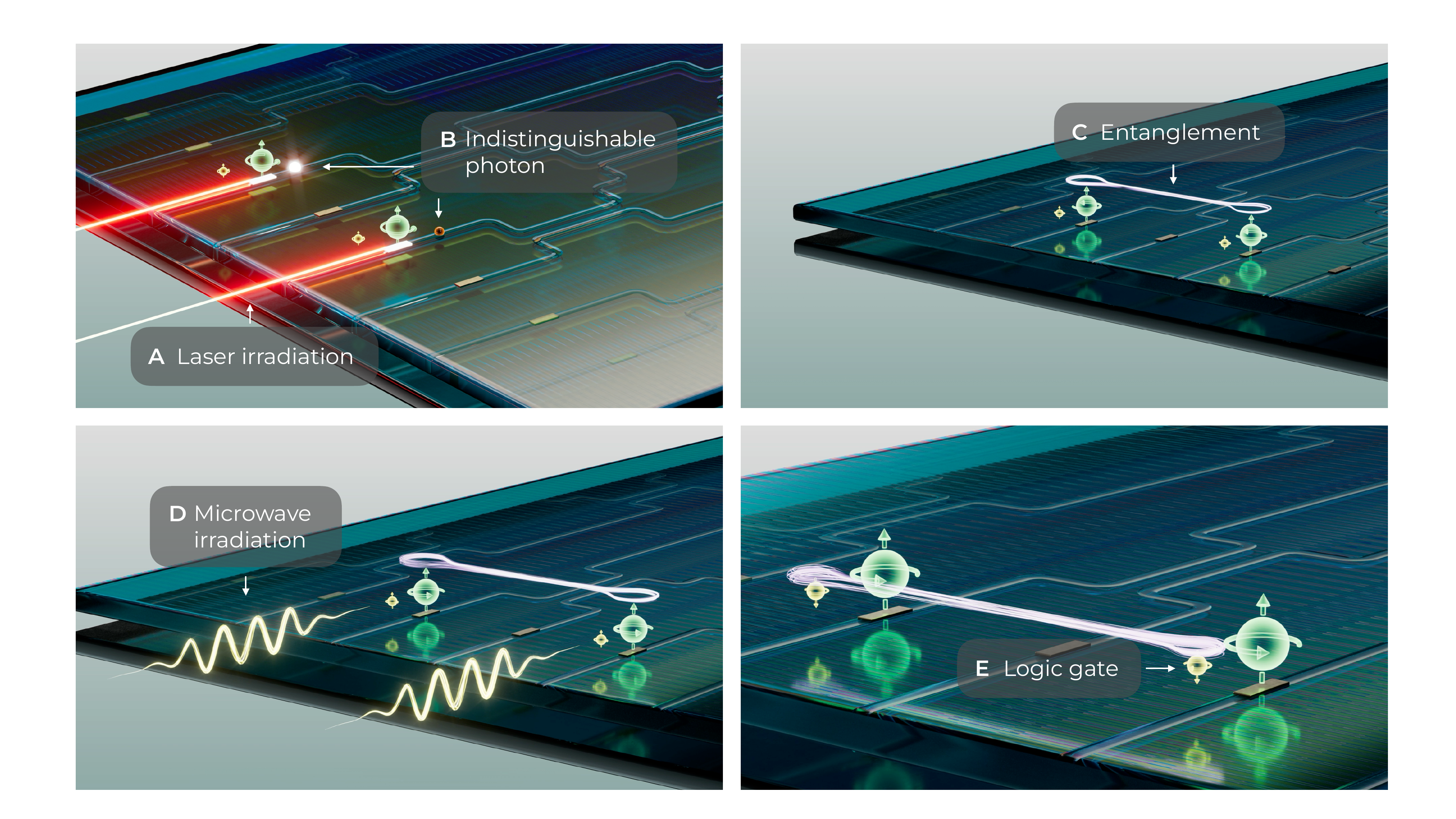}
\caption{The process for generating a unversal two-qubit gate. Heralded entanglement is achieved with the following sequence: Laser irradiation (A) excites two BiPhi molecules, emitting an indistinguishable photon (B), which is then detcted in the BSM station. A successful detection of two photons according to the Barrett-Kok protocol, leads to heralded entanglement (C) between the two electron spins. Local electron-nuclear gate is then implemented by a MW irradiation sequence (D), performing a CNOT between the nuclear memory and electron spin. Finally, optical measurement of the electron spins teleports the logical 2-qubit gate (E) to the nuclear memory spins.}
\label{fig:4}
\end{figure}

A universal two qubit gate, or a quantum error correction syndrome measurement is performed with the following steps:
\begin{enumerate}
    \item \textbf{Heralded remote entanglement} between the QPIs of two qubit cells. These can be performed in parallel between all pairs.
    \item \textbf{Local electron-nuclear gate} (\textit{e.g.}, CNOT) between the qubit and the QPI, adding the qubits to the entangled state.
    \item \textbf{Optical measurement of electron spins}, transferring (teleporting) the desired gate\cite{eisertOptimalLocalImplementation2000a} or measurement syndrome to the qubits of the two qubit cells.
\end{enumerate}

\subsection{Heralded remote entanglement}
Remote entanglement between two qubit cells is generated by emitting an indistinguishable photon from the QPIs of the two qubit cells whose "which-path" information is erased \cite{plenioCavitylossinducedGenerationEntangled1999,boseProposalTeleportationAtomic1999a,cabrilloCreationEntangledStates1999a}. As an example, one can use the double-click protocol (DCP): (1) initialise the electron and apply a $\pi/2$ microwave pulse; (2) optically excite once and collect emitted photons; (3) apply an electron spin inversion and repeat the excitation; (4) route both photons to a BSM station and herald on coincidence \cite{barrettEfficientHighfidelityQuantum2005a, deutsch1996quantum, bennett1996purification}.

Crucially, this enables \textbf{heralded entanglement}, meaning that if the photons were not detected we can repeat the entanglement attempts until we get the required photon detection.
This means that unlike photonic-only architectures for quantum computing~\cite{aghaee2025scaling, psiquantum2025manufacturable}, photon loss does not result in decreased entanglement fidelity, but only increased time until a successful entanglement.
This enables us to work with up to 70\% photon loss ($\sim -5$\,dB). Of course, this tradeoff is reasonable if the total time for an attempt is short enough to enable numerous attempts within the allotted time for the entanglement generation.

Taking BiPhi parameters and assuming reasonable values for the Purcell enhancement from the optical resonator ($F_P=30$), MW driving (30\,MHz) and detection efficiency (0.4), the total  time for attempting to create one bell pair is $\tau_{\text{bell-attempt}} \lesssim 100$\,ns \cite{deboneThresholdsDistributedSurface2024,nvision_blueprint}. Given the detection efficiency, assuming the double-click protocol and assuming we want 99.6\% of the nodes to achieve a successful entanglement, we arrive at a total of $\sim 6 \, \mu$s for the heralded entanglement step.\cite{nvision_blueprint} The 0.4\% of qubit cells that fail to become entangled remain idle for a single QEC round and are remeasured in the following round, which only slightly increases the total number of rounds and should not significantly affect the threshold. 
If necessary, entanglement fidelity can be further enhanced by entanglement purification using a nuclear ancilla spin.~\cite{nickersonFreelyScalableQuantum2014, deboneThresholdsDistributedSurface2024}

\subsection{Local electron-nuclear gates}
As mentioned above, a key advantage of molecular qubits is the ability to engineer the local spin environment, deterministically incorporating the nuclear spin qubits at specific locations in the molecule, yielding repeatable hyperfine coupling to the electron spin in the few MHz range. Using state of the art MW sequences, numerical simulations achieve fast gate times of $\sim 1 \, \mu$s. As the coherence time of both the electron (under dynamical decoupling) and nuclear spins is at least 2\,000-fold larger than the gate time, 99.9\% electron-nuclear spin gate fidelity is achievable, crucial for fault-tolerant operation.\cite{nvision_nuclear_electron_gates,nvision_blueprint}

\subsection{Optical measurement and initialization of electron spins}
For ultrafast optical measurements we leverage the existing photonic switch layer to turn the beam splitters into direct switches, enabling using the same optimized waveguides for detection of photons from individual qubit cells.
We utilize two key properties of BiPhi, namely the large difference of 15\,GHz between the optical transitions corresponding to different spin states and the very narrow optical linewidth $<40$\,MHz.
Thus, by exciting an optical transition of a specific electron spin state, we have negligible photon emission for the other spin states, even in an optical resonator. Therefore, by collecting very few photons we can measure and initialize to a specific spin state within 60\,ns with $>99.9\%$ probability.\cite{nvision_blueprint}

Measurements, as well as link establishment, introduce noise to the nuclei. In our setting, however, dephasing is substantially mitigated because of the structure of the Hamiltonian, in which the eigenstates do not carry a time-averaged magnetic dipole. Consequently, this dephasing has only a very small effect. For an in-depth analysis of a comparable scenario where the drive cancels the dipole, see Ref.~\cite{cohen2017protecting}.

\section{Distributed QEC via Floquetification of High-Rate Codes}
\label{sec:distributed_qec}
\subsection{Why Floquet codes for a networked architecture}
Surface-code realizations of distributed quantum computing utilize 4-qubit GHZ states to implement stabilizer measurements \cite{deboneThresholdsDistributedSurface2024,nickersonFreelyScalableQuantum2014,singh2025modular}. The same procedure could be be extended to realize distributed qLDPC codes using larger GHZ states.

These GHZ states are generated by fusing two-qubit Bell pairs in a costly process that incurs a significant overhead in both rate and fidelity, ultimately resulting in a low threshold for the QEC code. Moreover, hook errors arising during the creation of the GHZ state further reduce the code distance.

One way to overcome this is to use Bell states to teleport gates between different nodes \cite{chandra2025distributed} and use these to realize the code. This, however, introduces an additional overhead from the ancillas, which again lowers the code threshold and also strongly depends on the measurement error.

Floquet codes are structurally better matched to this setting: they are built entirely from pairwise (weight-two) measurements, and each non-local $X \otimes X, Y \otimes Y,$ or $Z \otimes Z$ check consumes a single shared Bell pair \cite{deboneThresholdsDistributedSurface2024,jacobyStairwayCodesFloquetifying2026,hastingsDynamicallyGeneratedLogical2021,sutcliffeDistributedQuantumError2025a}. Thus, realizations of distributed Floquet codes avoid both the costly fusion process \cite{sutcliffeDistributedQuantumError2025a} and gate teleportation, resulting in more efficient codes with higher thresholds.

\subsection{Floquetification procedure of qLDPC codes}
While Floquet codes are promising for distributed computing, they are less attractive than the recently introduced quantum low-density parity-check (qLDPC) codes, which can offer high rates (ratios of logical to physical qubits) and large code distances, at the expense of requiring non-local syndrome measurements \cite{breuckmannQuantumLowDensityParityCheck2021,gottesman14,tremblay22,panteleevAsymptoticallyGoodQuantum2022,leverrier22a,bravyiHighthresholdLowoverheadFaulttolerant2024,cainShorsAlgorithmPossible2026,zhao2026towards,kasai2026breaking}.

This gap can be bridged by efficient Floquetification procedures which we aim to use. It was recently shown\cite{jacobyStairwayCodesFloquetifying2026} that  Abelian two-block group algebra (2BGA) codes, a class that includes bivariate bicycle (BB) codes, can be Floquetufied in a systematic procedure while maintaining the efficiency of the original code. The procedure works by representing the static code's syndrome extraction circuit as a foliated ZX-calculus network in ($w-1$) spatial dimensions (where $w$ is the stabilizer weight), rotating the time axis, and decomposing each weight-$w$ stabilizer spider into a periodic sequence of pairwise measurements.
The resulting Stairway Floquet code preserves the parent code's encoding rate $r = k/n$ (since no ancilla qubits are needed for the syndrome extraction circuit), its detector structure, and its logical operators. Key properties are:
\begin{itemize}
    \item \textbf{Weight-two measurements only:} each check is a two-body Pauli measurement, directly implemented by one shared Bell pair.
    \item \textbf{No ancilla overhead:} the net encoding rate equals that of the parent static code.
    \item \textbf{Generality:} the method applies to any Abelian 2BGA code, including BB codes and quasi-cyclic lifted product (LP) codes, whose group structure is a product of cyclic groups\cite{breuckmannQuantumLowDensityParityCheck2021,panteleevDegenerateQuantumLDPC2021,bravyiHighthresholdLowoverheadFaulttolerant2024}.
\end{itemize}

We will use similar procedures to Floquetify the higher rate codes recently introduced \cite{cainShorsAlgorithmPossible2026,zhao2026towards,kasai2026breaking} and build upon this.

\subsection{Non-local checks via Bell pairs}
A non-local weight-two check is implemented by: (1) generating a heralded Bell pair shared between the QPIs of the two qubit cells \cite{campbell2008measurement}; (2) possibly distilling it by generating additional Bell pairs \cite{nickersonFreelyScalableQuantum2014}; (3) performing, in each qubit cell, a local gate between the nuclear-spin qubit and the entangled electron spin; and (4) measuring the electron spins and forwarding the outcomes to the decoder. Each non-local check consumes exactly one Bell pair; network throughput therefore scales linearly with the number of non-local edges in the partitioned code graph. While the thresholds for this procedure were analyzed in \cite{deboneThresholdsDistributedSurface2024} and found to be around $10^{-3}$, using Floquet codes should increase the threshold to the $1\%$ regime, comparable to state-of-the-art stabilizer codes.

\subsection{PIQC syndrome cycle time}
The syndrome cycle time is an important parameter for a fault-tolerant architecture.
In the Stairway-Floquetified LP code, each full syndrome cycle consists of $\lceil w/2 \rceil \approx 4$ sub-rounds of pairwise measurements (for weight-$w \leq 7$ stabilisers).
Within each sub-round, each non-local check consumes one heralded Bell pair ($\tau_{\text{Bell}} \approx 6 \, \mu$s), followed by a local CNOT gate ($\sim 1 \, \mu$s) and measurement ($\ll 1 \, \mu$s).
Since non-local checks can be partially parallelised, each sub-round takes approximately $\tau_{\text{sub}} \approx 7\text{--}20 \, \mu$s depending on the degree of parallelism and the number of sequential non-local checks per QPU.\cite{nvision_blueprint}

We estimate the full syndrome cycle time as:
\begin{equation}
\tau_{\text{cycle}} \approx 4 \times \tau_{\text{sub}} \approx 30\text{--}80 \, \mu\text{s}.
\end{equation}
This is potentially faster than other blueprints for FTQC~\cite{bluvsteinFaulttolerantNeutralatomArchitecture2026,cainShorsAlgorithmPossible2026,tripier2026fault}, and can be further sped up by an order or magnitude by modifying the architecture to achieve the ``reaction limit'', where the measurement time ($<100\, \text{ns}$ in our case) is the rate-limiting factor.~\cite{fowler2012time}.

\section{Conclusion}
\label{sec:conclusion}
PIQC presents a coherent pathway to distributed fault-tolerant quantum computing by combining four innovations: (i) photo-active BiPhi carbene molecular qubits with proven single-molecule spin-photon interface capabilities; (ii) deterministic nuclear-spin register; (iii) silicon photonic integration enabling fast and scalable Bell-pair generation and measurements; and (iv) Stairway Floquetification of high-rate qLDPC codes, including the LP codes that enable cryptographically relevant resource estimates\cite{cainShorsAlgorithmPossible2026,babbushSecuringEllipticCurve2026}.

Even on a small scale, PIQC already has the potential to significantly enhance current quantum devices, \textit{e.g.} in quantum communication as the missing quantum repeater link, and in quantum computing clusters as the high-fidelity photonic interconnect for coupling between monolithic QPUs.

The remaining experimental milestones, including electron--nuclear gate demonstration, photonic integration, heralded entanglement between QPIs and numerical simulation of Floquetified LP codes at scale, define a clear path toward a full resource estimate and, ultimately, a fault-tolerant PIQC processor.

\newpage

\bibliographystyle{naturemag}
\bibliography{references}

\end{document}